\documentclass[12pt]{article}

\usepackage{graphicx,amsmath,amsfonts,amssymb,dcolumn,amsthm,euscript}

\def\ee{\eea}
\def\be{\bea}
 
\def\eq{\eeq}
\def\bq{\beq}

\def\beq{\begin{equation}}
\def\eeq{\end{equation}} 
\def\bea{\begin{eqnarray}}
 \def\eea{\end{eqnarray}}

 \def\nn{\nonumber}

\def\v{{\dot q}_i+f{{\eij q_j}\over{\vec q^2}}}

\begin{document}

\title{\textbf{The  Universe   with an   Effective Discrete~Time  (II)  \footnote{  Version  (I) was presented    at the 2016 Carg\`ese Summer Institute, June 13-25.} }}
\author{\textbf{Laurent ~Baulieu }\thanks{baulieu@lpthe.jussieu.fr}\\\\
\textit{{  LPTHE UPMC $-$   Sorbonne Universit\'es,}}\\
\textit{{    4 Place Jussieu,    75 005 Paris, France}}\\
\textit{{    and }}\\
\textit{{  Max-Planck-Institut f\"ur Gravitationsphysik}}\\
 \textit{{ Albert--Einstein-Institut, 14476 Golm, Germany}}}
\date{}
\maketitle

\begin{abstract}
The  mechanism for  triggering the  universe   inflation  and its  sharp exit could be  that at very early periods the   time  variable was   discrete   instead of  smooth, defining a new transplanckian time physical scale.
 Alternatively, and perhaps equivalently, it could be the consequence that the metrics of the early universe   was  a   strongly concentrated   gravitational coherent state with very high frequency oscillations,    allowing local pair creations by a  generalisation to gravity of the Schwinger mechanism,  perhaps by creation of black holes  of masses  superior  to the Planck scale.
The lattice spacing between two clicks in the discrete time picture corresponds to  the inverse frequency of the gravitational coherent state in the other picture. In both cases,  a much lower time than the Planck time might represent  a new fundamental scale, which possibly gives a new UV regularised  type of physics.   To make   a possible   rough  estimation of the pair production  probability, we  propose that  the  oscillating    coherent state   metrics that reproduces  this very early geometry minimises 
     the Einstein gravity action  coupled  to interacting  1-,2- and 3-forms. Part of our intuition relies on   a condensed matter  analogy with laser-induced superconductivity.
     An independent section suggests that the new physical time scale we introduce for the Markovian discret time of the pre-inflation epoch can be identified with the stochastic time one uses for standard quantisation  in the post-inflation limit. In the pre-inflation phase  its discreteness is  the observable driving force for the universe evolution, while in the post inflation phase it becomes unobservable and it can can effectively replaced by the continuous time coordinate of standard quantum field theory. 

\end{abstract}

\newpage
\section{Introduction}

 Unknown   quantum gravitational physics   exists       at an  age of the universe of  the order of magnitude of the Planck Time
$\tau \equiv \sqrt {  \frac  { \hbar  G_F}{c^5}}\sim   10^{-44}s$ and earlier. According to current estimations, its effects might  last till an age as late as~$\sim 10^{-32}s $ 
and    for  a good fraction of the epoch of the   cosmic inflation.


In the current      phenomenological inflation model,   the space expands with an   ultra-fast  exponential pace,  between a time 
that is often estimated to be   $ \sim 10^{-36}s$   till  a time     between  $ \sim10^{-35}$  and  $\sim10^{-32}s $. Such time scales are  very large when  counted in      Planck  units. 
The inflation wave propagates
  at a  speed  that  is significantly smaller than the speed of light, therefore   we can   observe at the boundary of our past light-cone only a fraction of  the  Universe history. When one builds a model for the inflation, a strong constraint   is the observed   isotropy of the thermal  fluctuations of the CMB.

After the end of inflation, and a very sharp decrease of the cosmological constant,   the    universe    expands at a much less accelerated rate  following   the   dynamics of  the  current  elementary particle standard model (or its possible variations) with a  very small value of the cosmological constant.
 In this regime, the   classical Einstein theory   describes well enough    the gravitational effects for the cosmic evolution and  one can understand  the quantum effects   by   using  the mathematically consistent standard model of    electroweak and strong interactions.  

Nowadays,  the measured  cosmological constant is very small. Before and during the inflation, its effective value was much bigger.  This     note   is about an  attempt to physically explain what may have induced  this   transition.   
We will propose   a possible microscopic explanation for  the huge fluctuations that triggered the inflation.

The obvious  difference between   very  early times  and the present time is that  intervals of  the order of (or shorter than) the  Planck  Time $\tau$ are unobservable in  the post-inflation universe,  the opposite of what must happen in the transplanckian universe.   Nothing ensures us that time could be described  by    a smooth and continuous variable in very early times. We will in fact suggest that   the  time evolution was perhaps   discrete, by steps of $\tau$ or another much smaller time, which would define a new quantised time scale. We will propose  that  this could be an explanation  for the triggering of  the inflation,    naturally giving   a   strong change of the vacuum energy  due to fluctuations, that is,  a most likely  irreversible  change in the value of the cosmological constant. In fact, we propose that the duration of elementary gravitational    phenomena for the period of the inflation is a very small fraction of the Planck time of order $10^{-15}$,  by analogy with the known   ratio~$\sim10^{-15}$ between  the typical time scale of a microscopic chemical reaction and  the length time   of a macroscopically observed chemical reaction. The value~$\sim 10^{-15}$ also  applies to the ratio between the duration of elementary reactions that occur when liquid water ices and the duration  of     the glaciation  of a macroscopic quantity of water. 
 Analogously, when one consider macroscopic ferromagnetism transitions, the estimated time of an elementary spin flip is $\sim 10^{-12}s$.  Therefore we propose the relevance of a discrete  time that  clicks by steps of $ \sim 10^{-15} \tau \sim 10^{-60}s $, a new  fundamental microscopic  time scale that we assume   for  microscopic reactions at the scale of the Planck time, and we consider $\tau$   as an already macroscopic time unit in the early phase of the universe.
We also propose that the cosmological constant  can serve as an order parameter that determines in which phase  the universe physics description stands, either with  an observable discrete  time behaviour  or with an effective continuous time that averages smoothly over a large number or iteration of the Markov steps.

  Perhaps equivalently, and  more realistically to reasonably hope to do concrete computations,  we could  keep a continuous smooth time, but  propose   that,  when the universe was very small, the  gravitational field metrics was a  highly concentrated   singularity  free coherent state with an ultra fast frequency $\sim 10^{60} Hz $,   filling  either the whole universe or just a localised region of it.  Such a field could be cause of the  huge vacuum fluctuations that one often expects to  redistribute the  energy in a variation of the curvature as well as in the   energy momentum tensor of non negligible amounts of matter. The mechanism can take the form of   generalised  Schwinger effects  (adapted to the case of a strong oscillating gravitational field) that would create  pairs with masses of order     the Planck mass, maybe under the form of black holes.  The fate of the latter is to decay eventually in ordinary elementary particles. The end of  inflation is then the     global evolution from  the former coherent state to  a weakened one  filled with more and more matter. The transition is irreversible  because it is accompanied by an increase of the size of the universe and a dilution factor  with matter creation, such that, altogether,  the   value of the cosmological constant decreases   rapidly  enough. To  describe microscopically what triggers    this evolution,    one can  perhaps  only  use   the microscopic content of  the  standard gravity coupled to forms and to the   standard model  of  the electroweak and strong interactions.
  
Both cases, either a genuine discrete time or a rapidly oscillating background gravitational coherent state,   could equally well explain the transition between a large and a very small value of the cosmological constant.
  
   An  important difference between our proposal and  other ones   in a static  DeSitter  space, (see for instance  \cite{polyakov}\cite{henneaux}),~is analogous to that  existing  between  pair creations by the Schwinger mechanism within a high frequency resonant cavity~\cite{brezin} and  within   a static condenser. It provides  a much more dynamical frequency dependent factor to trigger vacuum fluctuations. In fact, in any given QFT, one expects that any given coherent state that defines a vacuum has always a non-zero probability to decay into another coherent state, which defines a new vacuum, with a shower  of particles that  enforces energy conservation. The quantitative details  obviously depends  on  the  theory. To possibly  get some concrete realisations, we  will in fact propose an Einstein action coupled to a set of 1-, 2- and 3-forms, all these fields vibrating at the same or comparable  super-fast frequencies at early times, at least in a local part of space. Getting non-singular solutions is preferable because one expects that the small time scale that one uses can be seen as an ultraviolet cutoff. The computation of such solutions for the metrics is a priori a feasible task, using appropriate boundary conditions. For instance, having an early  space with  a toroidal structure $S_3$ may facilitate the existence of time oscillating solutions. It is not yet obvious that we can have relevant oscillating solutions   when the universe has  strong spatial symmetries. There are interesting indications  that  non standard geometries occur   for certain couplings of gravity to gauge theories,    as for instance  in \cite{linden} and references therein. There are abundant descriptions of time dependent geometries in the literature but it is still unclear if even one of them is appropriate for   favouring  the  scenario presented here.  As we just say,  the hypothesis of a discrete time provides a physical scale for an absolute ultra-violet cutoff for all generality solutions,  thus one needs to exhibit {\it non-singular } solutions of general relativity to mimics the effect of a discrete time. Physically,  one can think to such solutions as sorts of gravitational laser beams,  which can be created  spontaneously because of    the non -linearity of general relativity  equations coupled to some matter fields.
   
  
  When solving the geometry of the space,  solutions that  exhibit  some aspects of the DeSitter space  are desirable 
   but they must be time dependent.  Moreover,   some of their  perturbative  excitations  around their   primary oscillations  must have a  long   range propagation in  the part of their future light cone,   as one 
   observes in the measured CMB.
    The duration  when one has strong enough   oscillating gravitational fields  that   give a big enough number of  pair creations   can  be very short,   but  the   frequency of the oscillations should be  extremely high, giving a new  time scale much smaller than the Planck time.
   and thus an energy  scale  much higher  than the Planck mass. 
    Our simplifying hypothesis is that that the coherent oscillations are fast enough   to  produce pairs of black holes, with masses      of the order of magnitude of  the Planck mass, maybe  rescaled by a scale factor~$\sim 10^{15}$, which relates    microscopic effects  to macroscopic ones,  that is,  masses of order~$ \sim 10^{35} Gev$.
  
  The mechanism is neither  in contradiction with the phenomenological explanation of the CMB by the amplification of perturbative quantum fluctuations  within a DeSitter background nor with  the reheating phenomena when particles are created in abundance,  The   coupling of a 3-form gauge field to gravity with Chern--Simons couplings involving    propagating $1$- and $2$-form gauge fields is perhaps an enhancement factor for getting such time dependent early universe backgrounds,  with appropriate time-depending geometries.  It  might be even  sufficient to have a local and big enough bubble of the early universe with such an oscillating behaviour, since   we can only  observe a fraction of all past events, the rest being outside of our past light cone.

Let us also contemplate some of 
our  present experimental knowledge about time.
Our best atomic clocks work at the scale of   $QED$
processes for atoms and allow us to detect time intervals down to
$~ 10^{-18}s$. This is the present smallest time scale for which one has directly verified the hypothesis of a smooth  and continuous time variable.
Having a continuous time down to the Planck scale is  not in contradiction with all  current astrophysics experiments.  
The LHC, our strongest microscope, uses elementary particle tools
instead of atomic tools. Using the tested  Lorentz invariance, it gives us a test for
shorter smooth time intervals that correspond to form factors with a
scale $~10^{-17 }  cm \leftrightarrow10^{-27s}   $.  This accuracy is completely out of scale as compared to the smallness of the Planck time.
In fact,  testing the hypothesises
 of continuous smooth time and  space variables are
quite different subjects, although they appear as naturally related, assuming Lorentz invariance. 
 The latter has been well verified at  the  LHC scale, but it can alway be questioned at much smaller space and time scales, which is a justification for our  proposal of a new time scale, much smaller than the Planck time. 

 Therefore, given the minimal scales of space and time that have been experimentally reached, there is  no contradiction for having  theories with a smooth time variable in our epoch and a  discrete discrete one  in the early ages with a lattice spacing quite small in Planck units, in a regime for which we don't have  yet a theory.
 Such a new microscopic time scale  can also play the role of a physical ultraviolet cutoff, in a way that is possibly compatible with the hypothesis of asymptotic safety of   S.~Weinberg.
  
A   possibly  fundamental discreteness of time  or   a  more conventional driving effect by  a huge gravitational coherent state is allowed by the smallness of the  very early universe and  its enormous gravitational  concentration and perhaps justified by the existence of the inflation.

In our scenario, it is not really  important whether the time of inflation was before of after  a date of order of the Planck time.  The current ideology is that the inflation occurred after the Planck time. There is some fuzziness about the moments when the Markovian  time we plea  for  could be {\it almost} approximated  by a continuous time. This is an analogous difficulty as for describing what really   often during some  intermediary moments of a phase transition,   for instance in ferromagnetism transitions that last for hours before reaching the magnetised form, while the duration of a spin flip is about $10^{-12}s$.  
 
  In fact, both scenario predict   driven quantum fluctuations, materialised by  avalanches of pair creations at high frequency that generalises the QED Schwinger mechanism. The   description with no discrete time  but with    the embedding of the system in an oscillating vacuum  is  more conventional. It may  justify in an easier  way  the  non-uniformity of the CMB by  the propagation of   quantum fluctuations around this quasi-classical gravitational field from one phase to the other. The  non-uniformity of the CBM  can take various aspects, depending on the initial conditions and on the   details of  the  propagation of microscopic transitions in a phase transition. Detailed computations of the propagation of these fluctuations for the transitions are maybe possible and can be very complicated, although one can do strong simplifying hypothesis, as one often does in condensed matter. On the other hand the scenario with a discrete time gets rid in an apparently easy way  of the difficulties  of singularities of classical and quantum field theories by  providing from the beginning an physical  ultraviolet cutoff. 
  
In the present    phase of the universe, such small time scales  cannot be directly measured. We will in fact  speculate that even if the Markovian time    is still running but is unobservable, it  can be mathematically identified  with  the stochastic time of stochastic time quantisation that is perfectly well suited to describe   the present quantum effects. We suggest that the latter      average  over a huge number of very smooth Markov processes. 
   
    To try to illustrate the effect of a discrete  time, we  first describe  a simple condensed matter example, where a laser beam possibly  triggers superconductivity for a sample of matter where valence bands and conducting bands are adequately organised  \footnote{More fancy mechanisms can be found in the literature for explaining high $T_c$ transitions for superconductivity.}. Then we will try to establish a dictionary to extend the idea for the effect of a discrete time on the evolution of the Universe at early times.
    
    Notice that the effects  of non-linearity   are surely an important ingredient  for  triggering strong transitions of the vacuum.

The non-linearity of the gravity theory makes very non-trivial the study of its coherent states, which should exist anyway, since we face a theory  with   infinite range.  Since we don't know the quantum version of gravity, it is   not possible to give the  precise  definition of its coherent states, except that they must be   the quantum states that are as near possible from classical solutions while being possibly subject to quantum fluctuations, following the ideas of Shr\"odinger.  This  non-linearity might be the cause  of the   highly concentrated and rapidly oscillating  gravitational fields that are  maybe   at the origin of  the huge  fluctuations  that modify    the vacuum   and thus  the value of the cosmological constant. 

 In the condensed matter example~\cite{Chamon}  that we    will describe as an   inspiration tool,  there  is also some non-linearity. As compared to the gravity case, the situation  is   somehow reversed. The coherent state  of  a laser beam is a solution of the  linear  Maxwell theory and   the non-linearity lies
in  the (phenomenological) shape of the band structures of the irradiated matter sample, as well as in  the  laser apparatus itself.  It is  also   known that  so-called topological defects play an important role for the description of some phase transitions. This was shown in  \cite{toulouse}, with  predictions  that were checked experimentally  by the discovery of new phenomenon in the hydrodynamic of some phases of Helium. The possible role
of  string theory defects   has  been widely discussed  for early time cosmology,   following the work of Witten   \cite{cosmicstring}. 

The last    section   (that was not presented in an earlier version of this paper)  concerns the idea we briefly mentioned above,   that the new physical time scale $\Delta T  \ll \tau_{Planck}$ we introduce for the Markovian discret time of the  inflation epoch can be identified with the stochastic time one uses for standard quantisation. It becomes physically unobservable (not to say confined)  in the post-inflation limit, by definition of having a  "phase transition" that freezes eventually  the effects of quantum gravity. We don't discuss here  possible   effects that the new microscopic time scale $\ll \tau_{Planck}$ 
 could help to uncover  about Weinberg ideas of asymptotic safety and about the necessity   of having a minimal length in string theory.

\section{Discrete time versus driven coherent state microscopic effects, the condensed matter example}

To understand the possible effects  of a theory with a discrete time, we may rely on condensed
matter considerations, where a laser beam, which is a powerful  coherent state with a given periodicity $1 /  2 \pi \omega$, can be used as "a discrete time
reservoir", or 
`` as a buffer for   large energy
fluctuations" to impose
its pace to a system. Such a buffer is more subtle than an ordinary
heat   and/or  particle bath. 
One then escapes the constraints of having   energy scales  that are governed only by thermal energy  $kT$ and  chemical potentials  $\mu$ in standard systems,   when the  standard Fermi--Bose  distributions  make so difficult the prediction of high temperature superconductivity.
In fact, the possible effects of   laser irradiation in
condensed matter were foreseen as early as in the 70's  in \cite{russians}.

Condensed matter   models with such a  phenomenological periodic discrete time implying  transitions  have been  recently considered. They involve   the irradiation by  a strong  enough  external laser beam over   a well structured  sample of matter with some conduction and valence bands. The whole apparatus is    thermalised   with an  elementary heat bath that controls temperature. Having no irradiation, the standard Fermi-Bose distributions  are actually rather restrictive for predicting strong transitions, especially superconductivity at not so low temperatures.  On the other hand, an external intense  laser beam  with frequency $\omega$ introduces another energy scale $ \hbar \omega$ and can  simulate a driven discrete time behaviour by steps of $ 1/  2\pi\omega$.  A refined  shape for the periodic  pulses of the laser can   make the situation even  more interesting, by introducing other  time scales  and thus other energy  scales. The whole system simulates macroscopically  an effective non-conservation of the energy, although
 energy is conserved at the microscopic level when one computes all  the photon exchanges between the laser and the matter.

 In theses examples, a given order parameter for the irradiated matter can drastically change because of predictable macroscopic fluctuations in the energy.   As foreseen in \cite{russians}, one can  invent and possibly trigger    new exotic phase transitions,  because, as already said,  one avoids  the    constraints of genuine  Fermi or Dirac statistics.   
 For instance, the  resistivity of an irradiated sample can change sharply, as if there was a phase transition. 
 When time is   discrete, one basically  escape  the   difficulty of getting stable Cooper pairs at high temperature because  the observed   energy is  effectively not conserved between two successive times :  the measured   energy  of the sample can jump by possibly large quantised and  uncorrelated  gaps.   Thus     non trivial       fluctuations  can  emerge that can drastically change the aspect of the irradiated   domain, for instance by putting in a stationary way all electrons in the conducting band.
  Then, the standard temperature energy scale $kT$ just becomes an external  parameter. The energy quanta  that allows the huge transitions  is  $\hbar \omega$ rather than $kT$, where  $\omega $ is  the frequency of the coherent beam.    One can  thus theoretically    stabilise e.g. Cooper pairs at  rather high values of the temperature.   It is thus suggestive to   call  an   irradiating beam as  a ``discrete time reservoir",  giving  a  sort  for equivalence principle between a driven period force by irradiation by a coherent state and genuine
discrete time effects.

Of course,  in condensed matter,  the discrete time is  just a formal  but convenient mathematical trick for a macroscopic description within   a statistical model. We know in this case that  the   underlying microscopic physics  is the well mastered quantum electrodynamics with a continuous time. The behaviour of   order parameters such as the conductivity     is driven  by the effect of this  effective discrete time intervals  cadenced by the   frequency $\omega$ of the  external laser (or the period between its pulses, or, even more subtlety by  some combination between the pulse width  and their periods). The   energy conservation is globally   ensured by the   power consumption of the laser  that balances the property that  (almost) arbitrary numbers of photons of the same energy   $\hbar \omega$  are   microscopically  continuously pumped in and out of the laser beam, ensuring the macroscopically observed effects of transitions due to the fluctuations that are inherent to a theory with a discrete time. This is   the buffer strength of the laser that drives the effective       discrete  time  behaviour.

 For the   sample that is irradiated  by the laser beam, however, everything  happens as if time is  discrete,  clicking by intervals  of $1/2\pi \omega $. Then, between two clicks,   energy is conserved only modulo integer numbers of quantas of energy $  \delta E_ {\rm {elementary}}= \hbar\omega$. So, between two successive times  $n/ 2\pi  \omega $ and $(n+1) /2\pi \omega $, the observer will measure  that energy is conserved only modulo integer values of $ \delta E_ {\rm elementary}=\hbar\omega$. In other terms, at any     given  possible observation time, the measured energy of the sample can be
\bq
E  \sim E \pm N   \delta E_ {\rm elementary} \ ,
\eq
with no small bound on $N$.  As a result, the behaviour of the sample can   change and simulates a phase transition. For instance when it is traversed by an electric current, the Joule effect becomes  completely different, leading possibly   to a stimulated superconductivity, at any  reasonable 
 value of the temperature, possibly for values of the temperature $T$ much larger than in the   situation with no laser illumination.  Of course the effect stops   soon after one unplugs the laser, and the system comes back to its original state, as  a sort of crunch.  Notice that for the current that measures the effective resistivity, the existence of the laser beam is irrelevant, what  only counts  is the possible access to the appropriate conductivity band, independently of the microscopic mechanism  that triggers this access.  As for the observer who measures the resistivity, he can  be  considered  as blind to the existence of the laser. If it is so, he might  conclude to the existence of a discrete time.

The   theory with a discrete time implies   from first principles  that  the integer $N$ can be as large  as one wants. When one goes to the details,   the standard energy conservation is  recovered   for   not too big values of $N$  and 
$ \omega \to 0$ or $\hbar  \to 0$, or more generally, $\hbar\omega \to 0$.

The smart  theoretical case  elaborated in \cite{Chamon} 
  is when electrons are strongly and persistently expelled from a valence band to a conductivity band of a well built sample of matter, even for not so small temperatures, as long as the laser beam is  activated. The curve in Fig. 4 in \cite{Chamon}   shows the time evolution of the resistivity of the irradiated sample  with a  spectacular transition from a stable plateau to another stable one, with  a rather quick  evolution  that is either oscillatory  or damped, depending on the values of some parameters. The  details for the shape of the transition are in fact irrelevant, the overall effect of having a transition being the same.  This curve is very suggestive,
showing an evolution of an order parameter that is quite alike the one
one wishes for the evolution of the cosmological constant for the 
inflation epoch.

A relatively simple  example of condensed matter   with  driven  discrete time effects may thus lead   to  a model with very non-trivial  consequences and non trivial changes of an order parameter.  Notice also that  the observer might ignore the existence of the laser  and   explore the system only using the Floquet   theory about genuine  discrete times, which has  becomes popular in condensed matter.

We now pass to the gravity case.

\section{Extension to Gravity at early times}


How should we describe  time intervals and time evolution in the very early universe ? In this epoch, nothing ensures  us that we have a smooth behaviour for the time variable.  Should we  maintain the assumption   that energy conservation works in the early universe as it does   when  time translation invariance exists  with a smooth time variable because of the Noether theorem ? 

  Here, we will assume that, at  a very early time,  say  at least much before~$\sim  10^{-32} s$, and maybe even much before the Planck time, the   continuous  time evolution   must be replaced    by the evolution in function of  a discrete  time  coordinate, either fundamentally or effectively.  There is    the natural question whether Lorentz invariance is still present in this regime;  it may happen that space is then either    smooth   or  discrete to maintain its relativity with time.

We  assume that when gravity was very dense everywhere,    time was possibly discrete,  with a yet undetermined theory, but   some basic principles  of    quantum mechanics were still operating.  They are:  (i) the uncertainty principle between some   variables and their conjugates  ;  (ii) the existence of coherent states for making the bridge between classical and quantum fields.  These properties can be demonstrated in ordinary quantum field theory and quantum mechanics. For quantum gravity, with or without a discrete time, it is quite reasonable to assume that they remain true.

In the ordinary case of $QED$,  the way coherent states  can   drive the system by  microscopic interactions between  photons and  atoms    is  well understood. One can compute the rate  pair creations by the Schwinger effect in intense laser beams (at least two face to face laser beams are needed to preserve the Lorentz invariance of a pair creation). Basically, one has the evolution  between a given coherent state and another one, with a creation of matter.  We will in fact assume    a sort  of equivalence principle  in quantum gravity, between  the effect of a fundamental discrete time and a forced time evolution by a periodic strong  coherent state, a phenomenon that we suppose may have   occurred in the conditions of the  early Universe. The mass of the  created pairs  is assumed to be higher, and perhaps much higher, than  the Planck mass, possibly under the form of  black holes, whose further evolution remains to be thought of.

We thus suggest that simple predictions can be drawn, for the  triggering of inflation, without   doing too many   model-dependent hypotheses.  Eventually, one has to verify that   the   inflation effect  triggered by a discrete time behaviour  is  compatible with  the phenomenological   description  by the standard inflaton field, with an appropriate potential. From our point of view,  the  existence of an inflaton   field   is clearly not needed, although it gives a satisfying phenomenological description.

It is suggestive enough to state  that when (i) the percentage of dark energy is very high and  (ii) the size of the universe is very small, the enormously  concentrated dark energy is nothing but the energy  of a  gravitational  coherent state, which   determines  the size and the vibration of space. At this time, the universe   is thus analogous to    a resonant cavity. One can be less ambitious and     consider only  a  bubble of the whole space, where the relevant fluctuations will occur,  in particular the bubble  that will be the whole past as seen from the light-cone past of an observer in the present epoch.
  It can drive huge fluctuations of the vacuum of the gravitational and standard elementary particles model, leading to the eventual  production of ordinary and dark matter by the Schwinger effect, so that, effectively, the universe is driven  toward a phase with a  very small cosmological constant. Some of the vacuum energy of the gravitational coherent state is  in fact  transformed into the energy momentum tensor of ordinary and dark matter, initially under the form of decaying objects with masses of the order of the Planck mass,  in interaction with a thereby attenuated  coherent gravitational  state, given a significant decrease of the proportion of  dark energy.    One may ignore  the  exact  details that occur all along   the transition, which could  follow various     paths, analogous to those that occur  when, e.g., ice melts into water in a few minutes, while, in comparison  each step is done by  minuscule increments of time of the order of that for chemical reaction $\sim 10^{-15} s$. In the case of the   melting of the ice    or the icing of   liquid water, there are many possible paths for the propagation of the process. Eventually, the structure of ice becomes very simple, (by this I mean that its structure  is not  very far from that of a     crystal). So,  we may believe that the impressively simple  isotropic shape and causal structure of the   presently observed CMB and its  phenomenological description by the inflaton model 
  and a simple effective potential is one other wonderful  example that   nature  often choses   simplest mathematical tools for a macroscopic   description of systems with a much refined  microscopic structure.

Supposing that  the uncertainty principle  remains  true, one     predicts  that there will be ordinary    energy  quantum fluctuations of the order of the Planck mass  $M _{Planck} =  \sqrt {  \frac  { \hbar   c^5}   { \hbar  G_F}    }  \sim  
10^{19} $~GeV. However, they are    relatively extremely  small,   since  the universe   energy  is certainly vastly superior to the mass energy  of its nowadays existing  $\sim 10^{80}$~nucleons. It is  hard to expect non trivial   transitions simply from  such relatively small fluctuations, when one assumes that time is a smooth variable.    

The hypothesis of  having a discrete periodic  time    allows us to go further with not too many thinking. Effectively,  since there is no way to go to intermediate times between  $n \Delta T$ and $(n+1)  \Delta T$,   
the energy  is no  more a  conserved quantity. Rather,   as  a prediction of the    quantised Floquet theory,  when one measures the   energy at  these successive times,     it is   conserved, but only    modulo integers numbers time the Planck mass (if it is the scale for the discreetness of time), that is  
\bq
E^{(n)}_ {universe}   \to  E ^{(n+1)} _{universe}   = E^{(n)}_ {universe}  \pm {\cal  N}^{(n)}\    M  _{Planck} 
\eq

The value of  the integer $ {\cal  N}^{(n)} $ can be  big enough for a given value of $n$. When one applies this to a gravity theory with a discrete time  variable, one sees that  there is always the possibility of the needed vacuum   energy fluctuation   for triggering inflation.   The needed time can be just a few units of the elementary time quanta or much longer. At very early times the time quanta we propose is 
$\Delta T\sim 10^{-15}\tau \sim 10^{-60}s$. In fact, after reaching a long enough time $(n+1)  \Delta T$, the vacuum may  be a state where the time can be described as a continuous  variable, and the  observed  post-inflation regime using the standard model description  can start.
 The proportion between  the relative percentages of dark energy and   matter energy,   and  thus the value of   the cosmological constant    will be changed    after the transition. In our scheme most of the matter should appear by a generalised Schwinger effect in an oscillating gravity field. 
 When the universe expands and becomes  filled with  more matter, dark energy gets diluted and it looses   its  coherence,   so its influence on matter become much less important, apart from the  classical  gravitational attraction.
 
When the   time  becomes effectively a smooth variable, intervals as short as $\Delta T$  become unobservable,  the  elementary  processes   last for much longer time than  $\Delta T$  and the universe follows its slow evolution, as it is  now, with a smooth time formulation. A reversing  crunch looks hard to think of, once   dark energy become more and more diluted,  since the probability of fluctuations becomes smaller and smaller.

\section{Estimating the gravitational Schwinger effect in the early Universe}

We wish to be a bit more quantitative and try  to estimate the rate of   Planck mass pair production in a a strong  pure gravitational coherent state  with  very fast oscillations.


 Our quantitative evaluation of the basic mechanism  can only be  very heuristic.  
 
There has been numerous excellent  papers using  the DeSitter space for studying inflation scenarios   with a cosmological term, for instance \cite{polyakov}\cite{henneaux},  although such spaces are basically instable  and  nothing can really move inside them. It has been realised that they    may  have a tendency to create matter by the Schwinger effect. The  back-reaction of this phenomenon is a non-trivial subject. One  understands that any classical field distribution is subject to instability, as emphasised  decades ago by Pauli and Schwinger. However, we believe that  the Schwinger effect of a static field is a too smooth effect to induce a decisive phenomenon such as the beginning of inflation. Coming back to the condensed matter electromagnetic-induced transition for superconductivity, we also understand   that  a laser with an appropriate frequency is more likely  to trigger the phenomenon rather than a static  condenser that can only provide a constant electric field.


The formula and reasoning  of  the production of pairs in  time depending fields  is explained in  \cite{brezin}.
For  high frequency colliding  laser beams the probability of production of a pair in an electric field $\sim E \cos\omega t$     can be computed to be proportional to 
\be  (  \frac{eE}{2\omega mc})   ^ { \frac{4mc^2}{\omega}}.
\ee
  The Schwinger  formula for the probability of pair creation in a constant static electric field  $E$ is
\be
P\sim  \exp
-\frac {m^2c^3}{
eE}.
\ee

Both formula  can be demonstrated  microscopically  within  $QED$ and one can interpolate between them. In the static case,  the   result  boils down   to the fact  that the probability for creating a pair  is  the exponential of the ratio of two relevant scales: the mass  of the threshold for creating a pair, and the work of the classical field on the Compton length of the particles to be pumped out of the vacuum. The formula must  by completed by
by some perturbative infinite resummation in the case of oscillating fields~\cite{brezin}. The phenomenon is often computed as a vacuum fluctuation using  the  coherent state formalism of $QED$.  

 Let us very crudely give a possible estimation of    the Schwinger  probability for  creating  a  pair of mass $m$ ($m$~is typically the Planck mass)  
 in a  Schwarzschild gravitational field far away  from the horizon. The physical arguments  is the following. In a sphere with mass M and radius R, the average value of the gravitational field is 
$
|g|\sim GM/{R^2}
$;  For creating a pair of particles with mass $m$, the work on a Compton length is thus  about 
$
\frac {\hbar}{mc} \frac{ GMm}{ R^2}$. This quantity must replace  $eE \frac {\hbar}{mc} $ in the  $QED$ Schwinger formula, so one gets:
\be
P\sim \exp - (  \frac {c R^2} { \hbar GM }  mc^2).
\ee
For $m\sim m_{Planck}$, this formula gives  
\be
P\sim \exp -(\frac{   R^2}{L^2_{Planck}}
\frac  {m_{Planck}}{M}).
\ee

We are not so much interested by the static  case. Rather, we prefer the option  that the early Universe works  as a resonant cavity with     time dependent   gravitational coherent states, whose frequency is much higher than the inverse Planck time. This corresponds  to a fast oscillating geometry in general relativity.
To get an estimate, we blankly extend the $QED$  frequency dependent formula. With the same notation for the  mass and radius of the universe, the gravitational analogous of the QED formula  for creating  matter pairs with masses $m$ of the order of the   Planck mass  in an oscillating gravitational field  can be reasonably assumed to be    
\be 
\Big(  \frac{eE}{2\omega mc}\Big)   ^ { \frac{4mc^2}{\omega}}   \to   
\Big(\frac   {GmM}{R^2}
	 \frac   {1}{\omega mc}  \Big)^ { \frac   {\omega_{Planck}}{\omega} }
	 =  
\Big(\frac   { M}{m_{Planck}}\frac{L^2_{Planck}  }{  { R^2}}
	 \frac   {\omega_{Planck}}{\omega  }   \Big)^{ \frac   {\omega_{Planck}}{\omega} }	 
\ee
 That is, in Planck units for masses and lengths
 \be P\sim \exp     { \frac   {\omega_{Planck}}{\omega} } 
 \Big (   \log \frac   {M}{R^2}
	+\log \frac   {\omega_{Planck}}{\omega}  \Big )  
	\ee
 When the frequency $\omega$ of the gravitational coherent state is big enough, 
 with
 \be
  \frac   {M}{R^2}
	>>   \frac  {\omega}  {\omega_{Planck}},
	\ee
  the probability of matter creation can be large enough for a time that is  long enough to create the matter whose amount will decrease substantially  the value of the vacuum energy at the requested rate, which depends on the value of $\omega$. The scenario of a stable oscillating gravitational field seems therefore  more appropriate  for pair creations than a  static one,  which  effect is very  soft  in comparison.  Doing more concrete computations would be extremely useful.

\section{Beyond the DeSitter case and the coupling to oscillating fields    as an enhancement factor for pair production }

The mechanism we invoke implies the existence of a fast frequency gravitational coherent state when the space was minuscule, which gives a new energy scale related to the frequency by the Planck constant, even if the standard $QFT$ is not applicable at such small scales of space-time.


Our hypothesis of a discrete  time,  (giving another scale than that of possible    horizons) could be more realistically supported if we had at our  disposal geometries with high frequencies time oscillations. For a simulation, one can think  of    replacing  the static singularity   of the   Schwarzschild--DeSitter  case  by  sources of  very rapidly oscillating black holes. They might  produce the needed high frequency gravitational 
coherent state for triggering the inflation,  and then  producing  particles. Knowing the difficulty to get not oversimplified    time dependent solutions of general relativity, we are in a difficult position to test the idea. In fact, we can only postulate  that certain general relativity  solutions exist that have the potential to     mimic the effect of a discrete time on the vacuum. A bonus would be that such oscillating solutions have no singularity.

   To get oscillating solutions, it might be necessary to couple the gravity theory to other theories that have themselves  natural oscillating solutions in flat space \footnote{ For instance   \cite{linden} and the references therein  show the existence of unexpected space oscillating solutions of coupled Einstein $SU(2)$-Yang--Mills equations  that are singularity free.}.

Here,  to  open the way to possible solutions, we wish to indicate  a generalisation of the  coupling of gravity to a  3-form gauge field, whose static effects related to the cosmological constant have been discussed extensively in the literature \cite{henneaux}. We find this possibility rather elegant    from a theoretical point of view, since it might be related  to string/branes arguments.

Such an abelian  field $A_3$ has a Maxwell type action, $\int F_4^*F_4,\ F_4=dA_3$  and accommodates a boundary term in 4 dimensions,
\be S(A,g)=
\int d^4x  \sqrt {-g}  ( \partial_{[\mu}  A_{\nu\rho\sigma]}\partial^{[\mu}  A^{\nu\rho\sigma]}) +\int dA_3
\ee
It couples to the gravitational field but it classically carries zero  particle degrees of freedom  in 4 dimensions.   Indeed, if nothing unexpected occurs, for instance an anomaly, the equations of motion and  the gauge invariance    imply that $F_4={\rm cte}\  d^4x$, consistently with the fact that gauge invariance  implies  that a 3-form has no physical degrees of freedom  in 4-dimensions and can only give an energy-momentum tensor proportional to $g_{\mu\nu}$.  However,  by doing    a  covariant    BRST quantisation, one  gets a Feynman type propagation of  the  unphysical degrees of freedom 
propagators that is compensated within on-shell amplitudes  by   that of the ghosts and ghosts of ghosts  degrees of freedom. So, the 3-form gauge fields   carries potentially some  dynamics that an anomaly can reveal, analogously to  what the conformal anomaly produces  in the Liouville 2d gravity. This makes the existence of a $3$-form gauge field even more  interesting. 

In  many   valuables papers, it is explained that the 3-form 
gives  a dynamical description for the existence of the static cosmological constant \cite{polyakov}\cite{henneaux}.  It has been  also  proposed  in  \cite{polchinski}  that the quantisation of the   four-form gauge flux  of $A_3$ can  make a variable contribution to the cosmological
constant.

 We can go  a little bit further.  Indeed, a ``latent"   anomaly of the gauge symmetry of the 3-form might enhance the possibility of  the wanted high frequency gravitational fields, possibly giving   a    time-depending value of the cosmological constant. Let us explain a possibility for such an anomaly.

 Anomalies may occur in all gauge symmetries of $p$-form gauge fields according to a simple algebraic classification \cite{GSB}. They break the  gauge symmetries of forms and the way one should count their  degrees of freedom, but they can be often compensated by introducing compensating fields, which restore the gauge invariance, but add new degrees of freedom. In our case, we have a 3-form in 4 dimension with curvature $F_4=dA_3+\ldots$. Then, a (mixed) anomaly can occur  if one can construct an invariant  6-form  $P_6(A_3, F_4, ....)$ in 6=4+2 dimensions, with the algebraic property  that $dP_6=0$ and $P_6$ is  proportional to the curvature $F_4$ of $A_3$, provided $F_4$ satisfies a Bianchi identity. Here the terms $....$ stand for fields that may couple to the $3$-form. To eliminate  this anomaly we need other fields to possibly compensate it by a generalisation of the Green--Schwarz mechanism, as in \cite{GSB}. Such fields can themselves
 develop anomalies, and we have to be a bit more systematic.

 Eventually  these various p-form gauge  fields will have physical degrees of freedom, and, once they are introduced to ensure that no anomaly can occur in the complete theory, they  will enrich the gravity content, and can possibly trigger new favourable oscillating  coupled classical solutions  beyond the known geometries,  at early times,  for making realistic our scenario.
 \def \v{{\small{ \wedge}}}
  \def \a{{\cal{ A}}}
  \def \f{{\cal{ F}}}
   \def \B{{\cal{ B}}}
      \def \G{{\cal{ G}}}
         \def \K{{\cal{ K}}}

So, besides the 3-form gauge field $A_3$, we introduce in a minimal scenario a 2-form  gauge field $\B_2$ and a  1-form gauge field $\a$ (or a collection of them, arranged as U(1) fields or as the elements of more complicated Lie algebra), all of them  
  with  curvatures   involving Chern--Simons couplings, as follows,
 \be \G_3 = dB_2+\a\v d\a\ee
 \be
\f =  d\a     , \ee
with the following Bianchi identities, which are non-trivial due to the mixed Chern--Simons coupling in $\G_3$
 \be d\G_3 = \f\v \f \ee\be
 d\f =  0  .\ee 
 These fields  get coupled to the 3-form $A_3$ for which we  can improve   the curvature as   \be F_4 =dA_3 \to F_4=dA_3   +   \a\v \G_3  
 \ee
 \be dF_4=0 \to  dF_4=\f\v\G_3.\ee
 By inspection, we find only one possible invariant 6-form depending on the forms
 $A_3,\B_2,\a$. It is 
 \be P_6=\f \v  \f\v \f ,\ee  but it can me made trivial by a Green--Schwartz mechanism. Indeed,   
\be\f \v  \f\v \f =  d(\a\v \f\v \f) ,  
 \ee
so  the anomalous  $QFT$  effect of $\a\v \f\v \f$  can be canceled  thanks to the Chern--Simons coupling in $\G$. This gives  a local   Wess and Zumino counterterm  to be possibly added to the possibly anomalous gravity action that interacts with  the $p$-forms. Using descent equations in 4 dimensions,  it  can be deduced from the six-dimensional non-trivial identity:
 \be   \a\v \f\v \f  =d ( \B_2\v\f) -\G_3\v\f= -d(\a\v  \G_3 ).
 \ee
The possible mixed anomaly  compensating action is (equivalently)  either $\a\v  \G_3$ or    $ \B_2\v  \f_2$, with an appropriate coefficient.  This anomaly compensation mechanism must be triggered if the vacuum can produce pairs of chiral fermions, with a   quantised coefficient $\alpha$  for the Chern--Simons terms.
 Notice that  one must then change the definition of $F_4$ into   $ F_4=dA_3   +   \a\v \G_3$,   so the mechanism we were looking for  triggering a non trivial gravitational background is  subtle.
  In short, to prevent anomalies, we need   the presence of a U(1) gauge field $\a$ and a 2-form  $ \B_2$. Getting a series of forms $\a, \B_2,A_3$ coupled to gravity in 4 dimensions is rather natural, as remnants of an underlying possible and yet unknown geometrical argument. Eventually, the 2-form can be replaced by a scalar field, using duality, and there is no gravitational anomaly because we are in four dimensions. 
 

 So we consider the action
 \be
 \label{action} I_{anomaly\ free}=
 \int d^4x  \sqrt {-g}  ( R+  F_{\mu\nu\rho\sigma} F^{\mu \nu\rho\sigma}+
 \G_{\mu\nu\rho} G^{\mu \nu\rho}
 +
 \f_{\mu\nu} \f^{\mu \nu}
 )
 +\beta \int dA_3 +\alpha\int \B_2\v \f,
 \nn\\
  \ee
 where $\alpha$ and $\beta$ are numerical coefficients.  Whatever is  the chiral fermion content of the vacuum, this theory is anomaly free by adequate tuning of the geometric coefficient $\alpha$.
 
  The role of the 3-form is to give effectively a cosmological constant  after its BRST invariant gauge-fixing ;  but it also couples locally to the  1-form  and 2-form, which  are themselves  self-coupled. In short, the system is that of   a propagating  system  of a 1-form and a  2-form, also coupled to gravity, whose modes can possibly  trigger the high frequency gravitational field  that is needed in our scenario for mimicking the effect of a discrete time, with a geometry that involves    a cosmological term because of the anomalous 3-form. Because of the Green--Schwartz mechanism, the value of the cosmological constant may vary. 
  

 Do we have a time oscillating  metrics that satisfies the coupled equations of motions stemming from 
  $I_{anomaly\ free}$? In our scenario, all fields are supposed to oscillate at some frequency, but the coupled 3-form coupling   gives perhaps additional horizons analogous to those which occur in  DeSitter space.  If one wishes a solution that oscillates naturally, one must maybe  assuming that the space has the topology of the torus $S_3$ and impose some periodic boundary conditions; but we can restrict  ourself to the case where the oscillations only occur within a bubble of the manifold, whose propagation toward the future  becomes the  observed  horizon  from where we are  nowadays. Moreover, as explained earlier, one wishes solutions that are non singular.

 Eventually a cascade phenomenon of pair creations with masses of the order of magnitude as $M_{Planck}$ might  be triggered with a rate as it was roughly estimated in the first part of this paper.
 
 \section {Markov (pre-inflation)  and Langevin (post-inflation)  processes for the Universe}
 \def\T{{\Delta T}}
 The way we see the pre-inflation epoch ressembles a Markov process. 
It evolves   by periodic discrete  steps of a given  time scale $\T$  (we  suggested  $\T \sim 10^{-15} \tau_{Planck} \sim  10^{-60} s$), with the possibilities of fluctuations that are basically independent of the previous step history. Although we ignore the details of the theory, we proposed  that fluctuations   are  analogous to  a  Schwinger effect in a powerful   and high frequency gravitational coherent state, (coupled to matter fields as e.g. in Eq.~(\ref{action})). The inflation can be possibly   triggered   by one  of these fluctuations whose effect is to generate black holes with masses of the order of the  Planck mass.
 
 It is   tempting enough to interpret the    successive times  $T, T+\T, T+2\T, \ ....$ as the theoretical time of a lattice computation.  For an  evolution over a very large number of Markov   processes, it  often makes sense to consider the continuous limit for certain phenomenon with physical time scale $t\gg {\cal N}_t\T$, with ${\cal N}_t$ extremely  large. We can go further. If the corresponding physics has a description in terms of a theory that is continuous in the time $t$ with differential equations involving the differential $dt$,  (in our case, this must occur after the end of the  inflation, when the cosmological constant became minuscule),  the   infinitesimal intervals of  physical time $dt$ one uses to physically define the standard  QFT path integrals also       correspond\ to a huge number of (ultra-smooth) Markov processes, with duration $dt ={\cal N}_{dt} \T$, with ${\cal N}_{t}    \gg{\cal N}_{dt} \sim \infty$. We thus have the hierarchy for time scales,
 \be\label{scales}  \T \ll dt\ll t.
 \ee
 In fact, we can heuristically say that $t$ is the physical continuous time  in Euclidean form that `{\it emerges}' from the Markov time $T$.
 
 We call $T$ the microscopic time and $t$ the physical emergent macroscopic time. The later can be the   Minkowski time of quantum field theory or that of the space-time of string theory.  Strictly speaking $t$ is Euclidean, and we must consider QFT that are compatible with an inverse Wick rotation to define eventually a Minkowski time. Some of the details of the  eventual effective smooth time   QFT (or    string theory) must be related to those of the Markov theory. A spectacular simplification is that in the (very early time)  Markov theory, one avoids by definition the  classical zero time singularities of gravity  coupled theories with a  continuous time, simply  because of the  periodicity of the early time Markov process. In a general Markov process, the nature of   (phase) transitions may  sometimes exhibits more subtleties that in continuous time transitions, for instance with a beginning, and then a back-up, so that inflation may start, but  then quickly abort, or become oscillating at a given rate.  But it can equally well evolve with a certain probability to the eventual phase where one has a theory driven by an effective smooth time. 
 

 For processes with a macroscopic  effective  time that is large when it is  compared to the microscopic scale $\T$, a Markov equation can be often approximated by a Langevin equation where the stochastic evolution is represented by a random noise $\eta(T,x)$, with
 \be\label{markov}
 \frac{\Delta \Phi( T, x)}{\T}= \frac{\delta I[\Phi]}{\delta \Phi (T, x)} +\eta( T,x)
 \ee
 When $\T$ is very small, the Langevin equation is continuous,
  \be\label{markov}
 \frac{\partial  \Phi( T, x)}{\partial T}= \frac{\delta I[\Phi]}{\delta \Phi (T, x)} +\eta( T,x)
 \ee
Here  $I$ is a local action of some fields $\Phi$ depending on effective Euclidean coordinates   $x= (\vec x,t)$. Lorentz invariance is not  necessarily ensured at early times but should also emerge at   large times.
 The simplest mathematical hypothesis is that when time is discrete, so does the space, and one has originally a hypercubic symmetry that eventually becomes the Lorentz symmetry. 
 
 In the simplest formulation the noise is Gaussian, which means
 \be
< \eta(T,x), \eta( T',x')> \sim \delta( T-T' )\delta( x-x' )
\ee
For instance, for the Lagrangian we mentioned, the drift force of the metrics is a combination of the metrics $g_{\mu\nu}$, the Riemann curvature $R_{\mu\nu}$ and the energy momentum tensor  $T_{\mu\nu}\sim F^2$ of appropriate  $p$-form gauge fields.  
To explain the triggering of inflation and simulate the effect of a discrete time,  we suggested  that  if, the $p$-form gauge fields  oscillate strongly  enough, then  the metrics can also oscillate, yielding     a non-singular   oscillating geometry that is  subjected to an abundant Schwinger effect, creating pairs of black holes. In this description    the matter creation   is  a quantum effect implied by the presence of the Gaussian noise.  As we tried to explain, this is a physically appealing description of  the triggering of the inflation transition that separates both  phases where the time is effectively discrete and where it can be approximated as a  continuous and smooth variable.

Equation~({\ref{markov})
 allows one to compute 5-dimensional Euclidian Green functions  in $x$ and $T$
\be
 <\Phi ( x_1,T_1), \ldots,  \Phi (T_n, x_n)>
\ee
by first inverting the Langevin equation to get the fields $\Phi$ as functions of  noises $\eta$
 and then using the definitions of the correlation functions  of the noises.

 This provides a set of  Euclidean correlation functions  depending also on $T$ with a wealth of information. 
 We are interested in processes with  time scales much bigger that the microscopic pace $\T$.   They depend on the emergent Minkowski  time $x^0 $ that will be obtained  by doing an inverse Wick rotation on $t$, when  the $T$ dependence must disappear by an appropriate limit. We thus assist to a subtle substitution   of the Markovian time $T$ into a continuous Minkowski $x^0$. 
 
A  prescription to get rid of the $T$ dependence  to define the correlation functions of the    present time   $4$-dimensional smooth quantum field theory  by computing those of the $5=4+1$ dimensional theory    {\it at equal T } and take   the {\it limit} $T\to \infty$, namely 
\be\label{def}
<\Phi ( x_1), \ldots,  \Phi ( x_n)> ^{continuous \ Euclidean\   times  \ t}
\equiv
lim_{T\to \infty} <\Phi ( x_1,T), \ldots,  \Phi ( x_n, T)> 
 \ee
Eventually, one does  Wick rotations generically denoted as  $ t\to i x^0$, and the correlators can be used to define $S$-matrix elements. But these  special   correlators in~(\ref{def}) are nothing else  than those one obtains  from the  standard  Euclidean path integral quantisation of the action $I$ with the equilibrium distribution $\exp -\frac{S}{\hbar}$.  In fact, the stochastic time formulation has been initiated in the 60's as a rigorous way to define the path integral in quantum mechanics, and its equivalence theorems  (\ref{def}) for standard quantum field theories  have been proven with various degrees of rigour, following the work of Parisi and Wu \cite{parisi} in the 80's.
 
 We thus claim after this line of reasoning that, in the smooth time  (post-inflation) regime, the microscopic time $T$ that clicks by steps of $\T$ can be identified as the time of stochastic quantisation, which  is in some interpretation  the computer time of the standard  lattice quantum field approximation for the action $I$ and  allows one to obtain the Euclidian Green functions of standard path integral quantisation.
 In our scenario, the differential $dt$ of the smooth effective time~$t$ of quantum field theory can be somehow considered as an  agglomerate  of truly `submicroscopic'   discrete steps $\T\sim 10^{-60}s$, according to Eq.~(\ref{scales}). In fact, $dt$ is the infinitesimal euclidian time elements for the elementary processes in the continuous time standard path integral.  In the continuous time  regime,    the steps $\T$  are relatively so small  that they are completely invisible and cannot produce observable fluctuations. $dt$ is a sort of averaging of the $\T$ over a huge number of relatively smooth iterations. The  approximation where one averages over them breaks down in the very early time of the universe, when $t \to 0$, or, more precisely, when  one has  physically relevant intervals $t'- t$  of the order of magnitude of $\T\ll \tau_{Planck}
 $. But at this time, the Markov processes are most favourable to possibly trigger a  huge fluctuation that can precisely brings back    the universe to a regime where time can be considered as smooth, which physically  makes the scheme  a consistent one. 
  
When  the conditions are such that the  Markov processes are invisible, quantum field theory (or its maybe even smoother generalisation string field  theory) becomes an excellent and predictive  approximation of the world, simply because  the Markov processes are approximated    by continuous Langevin or Fokker--Planck processes that have a natural description in term of smooth quantum field theory when one averages on huge packages of    time steps $\T$. The Langevin equation determines the local action of standard QFT, its  drift force gives the  equations of motion and its noise drives the quantum fluctuations that bring eventually  the universe in the smooth regime.  There might  be a correspondence between the locality of interactions and the fact that Markov processes involve  only  neighbour to neighbour interactions. We suggest  that  the physical picture  of a discrete time that undergoes a transition toward a continuous observable time    justifies the prescription of selecting observables in the stochastic quantisation as the equal stochastic time correlations functions  of the complete theory, when the stochastic time is taken as infinity.  As a matter of fact, all my previous attempts to find a symmetry principle for this prescription   always failed.  Here one finds an unexpected physical argument to justifies the equal  stochastic time prescription to extract from a 5-dimensional theory (including the stochastic time) the standard 4-dimensional QFT's that are appropriate to describe the elementary particles in the smooth time limit.

 This approach involves  an ultra small    physical    time scale    that may  represents  a physical  ultraviolet cut-off  as was advocated for by S.~Weinberg in the quantum gravity dominated regime. Here, it  represents  the  submicroscopic scale of   Markov processes that are only observable in the phase of   the early universe.   Such Markov processes build a  more general approach than  the  standard path integral  where the stochastic time is only fictitious mathematical time.

 We are perfectly conscious of the speculative aspect of this idea  of  a new physical discrete time  building up   the continuous time of the low density universe,  with a hierarchy as in~Eq.(\ref{scales}).  In the     system of unities   $\tau_{Planck}=c=G_N=1$,  one understands that   the stochastic time steps define a natural ultraviolet cutoff for all physical objects of the continuous QFT approximation, including the strings. The transition between the epoch when the time variable must be considered as a periodic and discret (a~solid) and when it can be considered as smooth (a gaz or a liquid) 
 is in fact the end of the  inflation. Hopefully, this  new physical ultraviolet cutoff could  be used to solve some difficulties posed by the quadratic divergencies of phenomenological QFT's of elementary particles and some of  the hierarchy questions.
 
%

  \section{Summary}

Let us summarise  the rough  mechanism  we have imagined. When the Universe is at a scale of the order of the Planck length or maybe much smaller, it may  function as a resonant cavity for dark energy, that is, it is filled with   a non-singular oscillating gravitational  coherent state that defines its geometry. This state    oscillates at a   frequency of the order of magnitude of $\tau_{Planck}^{-1}  $ or much  higher. Such high frequency   powerful gravitational coherent states  can  trigger   locally  strong enough fluctuations    of   the vacuum,  which  may in turn  percolate into   a    transition  that reduces drastically   the  value of the cosmological constant with a probability in time that one can unfortunately only roughly estimate. This   microscopic scenario for the    beginning of inflation is not in contradiction with the  current phenomenological description of  the inflation with an effective inflaton field whose evolution is aimed to fit the inflation curve.  More research must be done for the existence of non singular  geometries with oscillating metrics~\footnote{While writing these notes, I became aware of  a  recent encouraging article  \cite{enrico} where   a simple time dependent  4D-geometry has been exhibited and  used for the creation of gravitons, and also \cite{calcagni}.}. We did some remarks to justify the possible (effective) existence of such an oscillating coherent state by a possible effective coupling of genuine gravity to 1,2,3-forms with Wess and Zumino terms to ensure that the theory is anomaly free.

  The   alternative hypothesis of a periodic discrete time   has the advantage  of providing us  an attracting theoretical 
  logics  for the existence of the strong  fluctuation that triggers inflation.  
 The use of a coherent state of dark energy actually mimics the effect of a discrete time in  a certain sense, since it gives a periodic behaviour to the vacuum energy momentum tensor.
  Condensed matter well chosen examples could  provide toy models to illustrate this analogy.

The last section  suggests that the  (physical)  Markovian discret time of the pre-inflation epoch can be perhaps identified with the stochastic time one uses for standard quantisation. Although it is still running in the post-inflation limit, it  becomes physically  unobservable for  all processes such as elementary particles scattering experiments   that have  gigantic time scales as compared to the duration of the time intervals of the   Markov process.  In the post-inflation regime the  effective time variable {\it emerges} as a right parameter for time ordering phenomena when one does the inverse Wick rotation of the Euclidean correlators  computed by solving the Langevin equations that approximate the Markov processes.

%
%
%
 

\vskip .5cm
\noindent { \bf Acknowledgements : } I thank all my colleagues with whom I discussed these issues. I am vey grateful for hospitality and support to  the Golm  Albert--Einstein-Institut where part of this work  was done.

\end{document}